\documentclass[english,aps,preprint,superscriptaddress]{revtex4-1}
\usepackage{lmodern}
\usepackage{lmodern}
\usepackage[T1]{fontenc}
\usepackage[utf8]{inputenc}
\usepackage{mathrsfs}
\usepackage{amsmath}
\usepackage{amsthm}
\usepackage{amssymb}
\usepackage{graphicx}
\usepackage{esint}

\makeatletter
\newenvironment{lyxlist}[1]
	{\begin{list}{}
		{\settowidth{\labelwidth}{#1}
		 \setlength{\leftmargin}{\labelwidth}
		 \addtolength{\leftmargin}{\labelsep}
		 }}
	{\end{list}}
\theoremstyle{plain}
\newtheorem{thm}{\protect\theoremname}
\theoremstyle{definition}
\newtheorem{example}[thm]{\protect\examplename}

\usepackage{amsthm}\usepackage{latexsym}\usepackage{bm}\usepackage{amsfonts}

\usepackage{babel}
\usepackage{enumitem}

\usepackage{hyperref}
\hypersetup{
	breaklinks = true,
    colorlinks = true,
    citecolor = {blue},
	urlcolor = {blue},
	linkcolor = {blue}
}

\makeatother

\usepackage{babel}
\providecommand{\examplename}{Example}
\providecommand{\theoremname}{Theorem}

\begin{document}
\title{Machine learning and serving of discrete field theories --- when
artificial intelligence meets the discrete universe}
\author{Hong Qin}
\email{hongqin@princeton.edu}

\affiliation{Plasma Physics Laboratory, Princeton University, Princeton, NJ 08543,
U.S.A}
\begin{abstract}
A method for machine learning and serving of discrete field theories
in physics is developed. The learning algorithm trains a discrete
field theory from a set of observational data on a spacetime lattice,
and the serving algorithm uses the learned discrete field theory to
predict new observations of the field for new boundary and initial
conditions. The approach to learn discrete field theories overcomes
the difficulties associated with learning continuous theories by artificial
intelligence. The serving algorithm of discrete field theories belongs
to the family of structure-preserving geometric algorithms, which
have been proven to be superior to the conventional algorithms based
on discretization of differential equations. The effectiveness of
the method and algorithms developed is demonstrated using the examples
of nonlinear oscillations and the Kepler problem. In particular, the
learning algorithm learns a discrete field theory from a set of data
of planetary orbits similar to what Kepler inherited from Tycho Brahe
in 1601, and the serving algorithm correctly predicts other planetary
orbits, including parabolic and hyperbolic escaping orbits, of the
solar system without learning or knowing Newton's laws of motion and
universal gravitation. The proposed algorithms are also applicable
when effects of special relativity and general relativity are important.
The illustrated advantages of discrete field theories relative to
continuous theories in terms of machine learning compatibility are
consistent with Bostrom's simulation hypothesis.
\end{abstract}
\keywords{Machine learning, Discrete field theory, Discrete universe}
\maketitle

\section{Introduction and statement of the problem \label{sec:Introduction}}

Data-driven methodology has attracted much attention recently in the
physics community. This is not surprising since one of the fundamental
objectives of physics is to deduce or discover the laws of physics
from observational data. The rapid development of artificial intelligence
technology begs the question of whether such deductions or discoveries
can be carried out algorithmically by computers.

In this paper, I propose a method for machine learning of discrete
field theories in physics from observational data. The method also
includes an effective algorithm to serve the discrete field theories
learned, in terms of predicting new observations.

Machine learning is not exactly a new concept in physics. In particular,
the connection between artificial neural networks and dynamical systems
has been noticed for decades \citep{Narendra1990,Narendra1992,Ramacher1993,HowseAH95,Wilde1993,E2017,Chen2018,Haber2018}.
What is the new contribution brought by the present study? Most current
applications of machine learning techniques in physics can be roughly
divided into the following categories. (i) Using neural networks to
model complex physical processes, such as plasma disruptions in magnetic
fusion devices \citep{Wroblewski1997,Vannucci1999,Yoshino2003,Kates-Harbeck2019},
effective Reynolds stress due to turbulence \citep{WuXiao2018}, coarse-grained
nonlinear effects \citep{Bar-Sinai15344}, and proper moment closure
schemes for fluid systems \citep{Han201909854}. (ii) Solving differential
equations in mathematical physics by approximating solutions with
neural networks \citep{Dissanayake1994,Meade1994a,Meade1994,Lagaris1998,Bailer98,Long2019}.
In particular, significant progress has been made in solving Schrödinger's
equation for many-body systems \citep{Carleo602,Nomura2017}. (iii)
Discovering unknown functions or undetermined parameters in governing
differential equations \citep{Bongard2007,Schmidt2009,Brunton3932,Rudy2017,Schaeffer2017,Baydin2017,Raissi2018,Cranmer2019,Gels2019,Raissi2019}.
As a specific example, methods of learning the Hamiltonian function
of a canonical symplectic Hamiltonian system were proposed in recent
months \citep{WuQin2019,Lutter2019,Bertalan2019,Greydanus2019,Zhong2019,Sanchez-Gonzalez2019,Chen2019,Toth2019}.
(iv) Using neural networks to generate sampling data in statistical
ensembles for calculating equilibrium properties of physical systems
\citep{Shanahan2018,Noe2019,Halverson2019,Cranmer2019a}.

The problem addressed in this paper belongs to a new category. The
method proposed learns a field theory from a given set of training
data consisting of observed values of a physical field at discrete
spacetime locations. The laws of physics are fundamentally expressed
in the form of field theories instead of differential equations. It
is thus more important to learn the underpinning field theories when
possible. Since field theories are in general simpler than the corresponding
differential equations, learning field theories is easier, which is
true for both human intelligence and artificial intelligence. Except
for the fundamental assumption that the observational data are governed
by field theories, the learning and serving algorithms proposed do
not assume any knowledge of the laws of physics, such as Newton's
law of motion and Schrödinger's equation. This is a stark contrast
to all other methodologies of machine learning in physics.

Without losing of generality, let's briefly review the basics of field
theories using the example of first-order field theory in the space
of $\mathbb{\mathrm{R}}^{n}$ for a scalar field $\psi$. A field
theory is specified by a Lagrangian density $L(\psi,\partial\psi/\partial x^{\alpha}),$
where $x^{\alpha}$ $(\alpha=1,...,n)$ are the coordinates for $\mathrm{R}^{n}$.
The theory requires that with the value of $\psi$ fixed at the boundary,
$\psi(x)$ varies with respect to $x$ in such a way that the action
of the system 
\begin{equation}
\mathcal{A}=\int L(\psi,\partial\psi/\partial x^{\alpha})d^{n}x
\end{equation}
is minimized. Such a requirement of minimization is equivalent to
the condition that the following Euler-Lagrange (EL) equation is satisfied
everywhere in $\mathbb{\mathrm{R}}^{n}$, 
\begin{equation}
EL(\psi)\equiv\sum_{\alpha=1}^{n}\frac{\partial}{\partial x^{\alpha}}\left(\frac{\partial L}{\partial\left(\partial\psi/\partial x^{\alpha}\right)}\right)-\frac{\partial L}{\partial\psi}=0\,.\label{eq:EL}
\end{equation}
The problem of machine learning of field theories can be stated as
follows:
\begin{quote}
\textbf{Problem Statement 1}. For a given set of observed values of
$\psi$ on a set of discrete points in $\mathrm{R}^{n},$ find the
Lagrangian density $L(\psi,\partial\psi/\partial x^{\alpha})$ as
a function of $\psi$ and $\partial\psi/\partial x^{\alpha}$, and
design an algorithm to predict new observations of $\psi$ from $L$.
\end{quote}
Now it is clear that learning the Lagrangian density $L(\psi,\partial\psi/\partial x^{\alpha})$
is easier than learning the EL equation (\ref{eq:EL}), which depends
on $\psi$ in a more complicated manner than $L$ does. For example,
the EL equation depends on second-order derivatives $\partial^{2}\psi/\partial x^{\alpha}\partial x^{\beta}$
and $L$ does not. However, learning $L$ from a given set of observed
values of $\psi$ is not an easy task either for two reasons. Suppose
that $L$ is modeled by a neural network. We need to train $L$ using
the EL equation, which requires the knowledge of $\partial^{2}\psi/\partial x^{\alpha}\partial x^{\beta}.$
For this purpose, we can set up another neural network for $\psi(x)$,
which needs to be trained simultaneously with $L$. This is obviously
a complicated situation. Alternatively, one may wish to calculate
$\partial^{2}\psi/\partial x^{\alpha}\partial x^{\beta}$ from the
training data. But it may not be possible to calculate them with desired
accuracy, depending on the nature of the training data. Secondly,
even if the optimized neural network for $L$ is known, serving the
learned field theory by solving the EL equation with a new set of
boundary conditions presents a new challenge. The first-order derivatives
$\partial\psi/\partial x^{\alpha}$ and second-order derivatives $\partial^{2}\psi/\partial x^{\alpha}\partial x^{\beta}$
are hidden inside the neural network for $L$, which is nonlinear
and possibly deep. Solving differential equations defined by neural
networks ventures into uncharted territory.

As will be shown in Sec.\,\ref{sec:Machine-learning}, reformulating
the problem in terms of discrete field theory overcomes both difficulties.
Problem Statement 1 will be replaced by Problem Statement 2 in Sec.\,\ref{sec:Machine-learning}.
To learn a discrete field theory, it suffices to learn a discrete
Lagrangian density $L_{d}$, a function with $n+1$ inputs, which
are the values of $\psi$ at $n+1$ adjacent spacetime locations.
The training of $L_{d}$ is straightforward. Learning serves the purpose
of serving, and the most effective way to serve a field theory with
long term accuracy and fidelity is by offering the discrete version
of the theory, as has been proven by the recent advances in structure-preserving
geometric algorithms \citep{Feng1985,Sanz-Serna1994,marsden1998multisymplectic,marsden2001discrete,hairer2006geometric,Qin2008VI-PRL,qin2009variational,li2011variational,Squire4748,squire2012geometric,squire2012gauge,Xiao2013,zhang2014canonicalization,zhou2014variational,he2015Hamiltonian,xiao2015explicit,xiao2015variational,ellison2015development,qin2016canonical,He16-092108,xiao2016explicit,zhang2016explicit,Wang2016,xiao2017local,burby2017finite,chen2017canonical,zhou2017explicit,He2017symplectic,Burby2017,kraus2017gempic,Xiao2018b,Ellison2018,Xiao2019Maxwell,Xiao2019,Xiao2019a,Glasser2019b,Shi2019,Xiao2019-comment}.
Therefore, learning a discrete field theory directly from the training
data and then serving it constitute an attractive approach for discovering
physical models by artificial intelligence.

It has long been theorized since Euclid's study on mirrors and optics
that as the most fundamental law of physics, all nature does is to
minimize certain actions \citep{Maupertuis1744,Maupertuis1746}. But
how does nature do that? The machine learning and serving algorithms
of discrete field theories proposed may provide a clue, when incorporating
the basic concept of the simulation hypothesis by Bostrom \citep{Bostrom03}.
The simulation hypothesis states that the physical universe is a computer
simulation, and it is being carefully examined by physicists as a
possible reality \citep{Beane2014,Glasser2019a,Glasser2019}. If the
hypothesis is true, then the spacetime is necessarily discrete. So
are the field theories in physics. It is then reasonable to suggest
that some machine learning and serving algorithms of discrete field
theories are what the discrete universe, i.e., the computer simulation,
runs to minimize the actions.

In Sec.\,\ref{sec:Machine-learning}, the learning and serving algorithms
of discrete field theories are developed. Two examples of learning
and predicting nonlinear oscillations in 1D are given in Sec.\,\ref{sec:Examples}
to demonstrate the method and algorithms. In Sec.\,\ref{sec:Kepler},
I apply the methodology to the Kepler problem. The learning algorithm
learns a discrete field theory from a set of observational data for
orbits of the Mercury, Venus, Earth, Mars, Ceres, and Jupiter, and
the serving algorithm correctly predicts other planetary orbits, including
the parabolic and hyperbolic escaping orbits, of the solar system.
It is worthwhile to emphasize that the serving and learning algorithms
do not know, learn, or use Newton's laws of motion and universal gravitation.
The discrete field theory directly connect the observational data
and new predictions. Newton's laws are not needed.

\section{Machine learning and serving of discrete field theories \label{sec:Machine-learning}}

In this section, I describe first the formalism of discrete field
theory on a spacetime lattice, and then the algorithm for learning
discrete field theories from training data and the serving algorithm
to predict new observations using the learned discrete field theories.
The connection between the serving algorithm and structure-preserving
geometric integration methods is highlighted.

To simplify the presentation and without losing generality, the theory
and algorithms are given for the example of a first-order scalar field
theory in $\mathrm{R}^{2}$. One of the dimension will be referred
to as time with coordinate $t$, and the other dimension space with
coordinate $x$. Generalizations to high-order theories and to tensor
fields or spinor fields are straightforward.

\begin{figure}[ptb]
\begin{centering}
\includegraphics[width=3in]{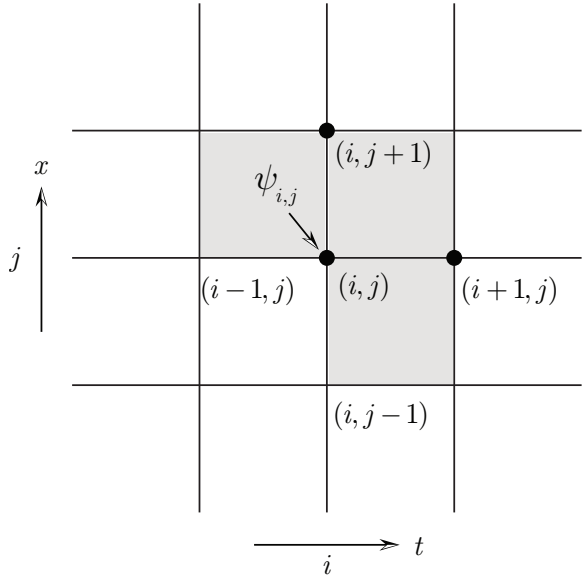}
\par\end{centering}
\caption{Spacetime lattice and discrete field $\psi$. The discrete Lagrangian
density $L_{d}(\psi_{i,j},\psi_{i+1,j},\psi_{i,j+1})$ of the grid
cell whose lower left vertex is at the grid point $(i,j)$ is chosen
to be a function of the values of the discrete field at the three
vertices marked by solid circles. The action $\mathscr{\mathcal{A}}_{d}$
of the system depends on $\psi_{i,j}$ through the discrete Lagrangian
densities of the three neighboring grid cells indicated by gray shading.}

\label{fig:lattice}
\end{figure}

For a discrete field theory in $\mathrm{R}^{2}$, the field $\psi_{i,j}$
is defined on a spacetime lattice labeled by two integer indices $(i,j)$.
For simplicity, let's adopt a rectangular lattice shown in Fig.\,\ref{fig:lattice}.
The first index $i$ identifies temporal grid points, and the second
index $j$ spacial grid points. The discrete action $\mathscr{\mathcal{A}}_{d}$
of the system is the summation of discrete Lagrangian densities over
all grid cells,
\begin{equation}
\mathscr{\mathcal{A}}_{d}=\Delta t\Delta x\sum_{i,j}L_{d}(\psi_{i,j},\psi_{i+1,j},\psi_{i,j+1})\,,\label{eq:Ad}
\end{equation}
where $\Delta t$ and $\Delta x$ are the grid sizes in time and space
respectively, and $L_{d}(\psi_{i,j},\psi_{i+1,j},\psi_{i,j+1})$ is
the discrete Lagrangian density of the grid cell whose lower left
vertex is at the grid point $(i,j)$. I have chosen $L_{d}$ to be
a function of $\psi_{i,j}$, $\psi_{i+1,j}$, and $\psi_{i,j+1}$
only, which is suitable for first-order field theories. For instance,
in the continuous theory for wave dynamics, the Lagrangian density
is 
\begin{equation}
L\left(\psi,\frac{\partial\psi}{\partial t},\frac{\partial\psi}{\partial x}\right)=\left(\frac{\partial\psi}{\partial t}\right)^{2}-\left(\frac{\partial\psi}{\partial x}\right)^{2}\,.\label{eq:Lwave}
\end{equation}
Its counterpart in the discrete theory can be written as
\begin{equation}
L_{d}(\psi_{i,j},\psi_{i+1,j},\psi_{i,j+1})=\left(\frac{\psi_{i+1,j}-\psi_{i,j}}{\Delta t}\right)^{2}-\left(\frac{\psi_{i,j+1}-\psi_{i,j}}{\Delta x}\right)^{2}\,.\label{eq:Ldwave}
\end{equation}
The discrete Lagrangian density $L_{d}$ defined in Eq.\,(\ref{eq:Ldwave})
can be viewed as an approximation of the continuous Lagrangian density
$L$ in Eq.\,(\ref{eq:Lwave}). But I prefer to take $L$$_{d}$
as an independent object that defines a discrete field theory.

For the discrete field theory, the condition of minimizing the discrete
action $\mathscr{\mathcal{A}}_{d}$ with respect to each $\psi_{i,j}$
demands 
\begin{align}
EL_{i,j}(\psi) & \equiv\frac{\partial\mathscr{\mathcal{A}}_{d}}{\partial\psi_{i,j}}=\frac{\partial}{\partial\psi_{i,j}}\left[L_{d}(\psi_{i-1,j},\psi_{i,j},\psi_{i-1,j+1})\right.\nonumber \\
 & \left.+L_{d}(\psi_{i,j},\psi_{i+1,j},\psi_{i,j+1})+L_{d}(\psi_{i,j-1},\psi_{i+1,j-1},\psi_{i,j})\right]=0\,.\label{eq:dEL}
\end{align}
 Equation (\ref{eq:dEL}) is called Discrete Euler-Lagrange (DEL)
equation for the obvious reason that its continuous counterpart is
the EL equation (\ref{eq:EL}) with $x^{1}=t$ and $x^{2}=x$. Following
the notation of the continuous theory, I also denote the left-hand-side
of the last equal sign in Eq.\,(\ref{eq:dEL}) by an operator $EL_{i,j}(\psi),$
which maps the discrete field $\psi_{i,j}$ into another discrete
field. The DEL equation is employed to solve for the discrete field
$\psi$ on the spacetime lattice when a discrete Lagrangian density
$L_{d}$ is prescribed. This has been the only usage of the DEL equation
in the literature so far \citep{marsden1998multisymplectic,marsden2001discrete,Qin2008VI-PRL,qin2009variational,li2011variational,Squire4748,squire2012geometric,squire2012gauge,Xiao2013,zhang2014canonicalization,zhou2014variational,he2015Hamiltonian,xiao2015explicit,xiao2015variational,ellison2015development,qin2016canonical,He16-092108,xiao2016explicit,zhang2016explicit,Wang2016,xiao2017local,burby2017finite,chen2017canonical,zhou2017explicit,He2017symplectic,Burby2017,kraus2017gempic,Xiao2018b,Ellison2018,Xiao2019Maxwell,Xiao2019,Xiao2019a,Glasser2019b,Shi2019,Xiao2019-comment}.
I will come back to this shortly.

For the problem posed in the present study, the discrete Lagrangian
density $L_{d}$ is unknown. It needs to be learned from the training
data. Specifically, in terms of the discrete field theory, the learning
problem discussed in Sec.\,\ref{sec:Introduction} becomes:
\begin{quote}
\textbf{Problem Statement 2}. For a given set of observed data $\overline{\psi}_{i,j}$
on a spacetime lattice, find the discrete Lagrangian density $L_{d}(\psi_{i,j},\psi_{i+1,j},\psi_{i,j+1})$
as a function of $\psi_{i,j}$, $\psi_{i+1,j}$, and $\psi_{i,j+1}$,
and design an algorithm to predict new observations of $\psi_{i,j}$
from $L_{d}$.
\end{quote}
Unlike the difficult situation described in Sec.\,\ref{sec:Introduction}
for learning a continuous field theory, learning a discrete field
theory is straightforward. The algorithm is obvious once the problem
is declared as in Problem Statement 2. We set up a function approximation
for $L_{d}$ with three inputs and one output using a neural network
or any other approximation scheme adequate for the problem under investigation.
The approximation is optimized by adjusting its free parameters to
minimize the loss function
\begin{equation}
F(\overline{\psi})=\frac{1}{IJ}\sum_{i=1}^{I-1}\sum_{j=1}^{J-1}EL_{i,j}(\overline{\psi})^{2}\label{eq:F}
\end{equation}
on the training data $\bar{\psi}$, where $I$ and $J$ are the total
number of grid points in time and space respectively. In Problem Statement
2 and the definition of loss function (\ref{eq:F}), it is implicitly
assumed that the training data are available over the entire spacetime
lattice. Notice that according to Eqs.\,(\ref{eq:dEL}) and (\ref{eq:F}),
first-order derivatives of $L_{d}$ with respect to all three arguments
are required to evaluate $F(\overline{\psi})$. Automatic differential
algorithms \citep{Baydin2017}, which have been widely adopted in
artificial neural networks, can be applied. To train the neural network
or other approximation for $L_{d},$ established methods, including
Newton's root searching algorithm and the Adam optimizer \citep{Kingma2014},
are available.

Once the discrete Lagrangian density $L_{d}$ is trained, the learned
discrete field theory is ready to be served to predict new observations.
After boundary conditions are specified, the DEL equation (\ref{eq:dEL})
is solved for the discrete field $\psi_{i,j}$. A first-order field
theory requires two boundary conditions in each dimension. As an illustrative
example, let's assume that $\psi_{0,j}$ and $\psi_{1,j}$ are specified
for all $j$s, corresponding to two initial conditions at $t=0,$
and $\psi_{i,0}$ and $\psi_{i,1}$ are specified for all $i$s, corresponding
to two boundary conditions at $x=0.$ Under these boundary and initial
conditions, the DEL equation (\ref{eq:dEL}) can be solved for field
$\psi_{i,j}$ for all $i$s and $j$s as follows.
\begin{lyxlist}{00.00.0000}
\item [{Step~~1)}] Start from the DEL equation at $(i,j)=(1,2),$ i.e.,
$EL_{1,2}(\psi)=0,$ which is an algebraic equation containing only
one unknown $\psi_{2,2}.$ Solve $EL_{1,2}(\psi)=0$ for $\psi_{2,2}$
using a root searching algorithm, e.g., Newton's algorithm.
\item [{Step~~2)}] Move to grid point $(i,j)=(1,3).$ Solve the DEL equation
$EL_{1,3}(\psi)=0$ for the only unknown $\psi_{2,3}.$
\item [{Step~~3)}] Repeat Step 2) with increasing value of $j$ to generate
solution $\psi_{2,j}$ for all $j$s.
\item [{Step~~4)}] Increase index $i$ to $2$. Apply the same procedure
in Step 3) for generating $\psi_{2,j}$ to generate $\psi_{3,j}$
for all $j$s.
\item [{Step~~5)}] Repeat Step 4) for $i=3,4,...,I$ to solve for all
$\psi_{i,j}$.
\end{lyxlist}
In a nutshell, the DEL equation at the grid cell labeled by $(i,j)$
(see Fig.~\ref{fig:lattice}) is solved as an algebraic equation
for $\psi_{i+1,j}$. This serving algorithm propagates the solution
from the initial and boundary conditions to the entire spacetime lattice.
It is exactly how the physical field propagates physically. According
to the simulation hypothesis, the algorithmic propagation and the
physical propagation are actually the same thing. When different types
of boundary and initial conditions are imposed, the algorithm needs
to be modified accordingly. But the basic strategy remains the same.
Specific cases will be addressed in future study.

The above algorithms in $\mathrm{R}^{2}$ can be straightforwardly
generalized to $\mathrm{R}^{n},$ where the discrete Lagrangian density
$L_{d}$ will be a function of $n+1$ variables, i.e, $\psi_{i_{1},i_{2},...,i_{n}}$,
$\psi_{i_{1}+1,i_{2},...,i_{n}}$, $\psi_{i_{1},i_{2}+1,...,i_{n}}$,......,
$\psi_{i_{1},i_{2},...,i_{n}+1}$. And in a similar way as in $\mathrm{R}^{2}$,
the serving algorithm solves for $\psi_{i_{1},i_{2},...,i_{n}}$by
propagating its values at the boundaries to the entire lattice. It
can also be easily generalized to vector fields or spinor fields,
as exemplified in Sec.\,\ref{sec:Kepler}.

It turns out this algorithm to serve the learned discrete field theory
is a variational integrator. The principle of variational integrators
is to discretize the action and Lagrangian density instead of the
associated EL equations. Methods and techniques of variational integrators
have been systematically developed in the past decade \citep{marsden1998multisymplectic,marsden2001discrete,Qin2008VI-PRL,qin2009variational,li2011variational,Squire4748,squire2012geometric,squire2012gauge,Xiao2013,zhang2014canonicalization,zhou2014variational,he2015Hamiltonian,xiao2015explicit,xiao2015variational,ellison2015development,qin2016canonical,He16-092108,xiao2016explicit,zhang2016explicit,Wang2016,xiao2017local,burby2017finite,chen2017canonical,zhou2017explicit,He2017symplectic,Burby2017,kraus2017gempic,Xiao2018b,Ellison2018,Xiao2019Maxwell,Xiao2019,Xiao2019a,Glasser2019b,Shi2019,Xiao2019-comment}.
The advantages of variational integrators over standard integration
schemes based on discretization of differential equations have been
amply demonstrated. For example, variational integrators in general
are symplectic or multi-symplectic \citep{marsden1998multisymplectic,marsden2001discrete,Qin2008VI-PRL,qin2009variational},
and as such are able to bound globally errors on energy and other
invariants of the system for all simulation time-steps. More sophisticated
discrete field theories have been designed to preserve other geometric
structures of physical systems, such as the gauge symmetry \citep{squire2012gauge,Glasser2019b}
and Poincaré symmetry \citep{Davoudi2012,Xiao2019Maxwell,Glasser2019,Glasser2019a}.
What proposed in this paper is to learn the discrete field theory
directly from observational data and then serve the learned discrete
field theory to predict new observations.

\section{Examples of learning and predicting nonlinear oscillations \label{sec:Examples}}

In this section, I use two examples of learning and predicting nonlinear
oscillations in 1D to demonstrate the effectiveness of the learning
and serving algorithms. In 1D, the discrete action reduces to the
summation of the discrete Lagrangian density $L_{d}$ over the time
grids,
\begin{equation}
\mathscr{\mathcal{A}}_{d}=\Delta t\sum_{i=0}^{I}L_{d}(\psi_{i},\psi_{i+1})\,.\label{eq:Ad1}
\end{equation}
 Here, $L_{d}(\psi_{i},\psi_{i+1})$ is a function of the field at
two adjacent time grid points. The DEL equation is simplified to 
\begin{equation}
EL_{i}(\psi)\equiv\frac{\partial L_{d}(\psi_{i-1},\psi_{i})}{\partial\psi_{i}}+\frac{\partial L_{d}(\psi_{i},\psi_{i+1})}{\partial\psi_{i}}=0\,.\label{eq:DEL1}
\end{equation}
The training data $\bar{\psi}_{i}$ $(i=0,...,I)$ form a time sequence,
and the loss function on a data set $\psi$ is
\[
F(\psi)=\frac{1}{I}\sum_{i=1}^{I-1}EL_{i}(\psi)^{2}\,,
\]
After learning $L_{d}$, the serving algorithm will predict a new
time sequence for every two initial conditions $\psi_{0}$ and $\psi_{1}$.
Note that Eq.\,(\ref{eq:DEL1}) is an algebraic equation for $\psi_{i+1}$
when $\psi_{i-1}$ and $\psi_{i}$ are known. It is an implicit two-step
algorithm from the viewpoint of numerical methods for ordinary differential
equations. It can be proven \citep{marsden2001discrete,Qin2008VI-PRL,qin2009variational}
that the algorithm exactly preserves a symplectic structure defined
by 
\begin{gather}
\Omega(\psi_{i},\psi_{i+1})=d\theta\,,\\
\theta=\frac{\partial L_{d}(\psi_{i},\psi_{i+1})}{\partial\psi_{i+1}}d\psi_{i+1}\,.
\end{gather}
The algorithm is thus a symplectic integrator, which is able to bound
globally the numerical error on energy for all simulation time-steps.
Compared with standard integrators which do not possess structure-preserving
properties, such as the Runge-Kutta method, variational integrators
deliver much improved long-term accuracy and fidelity.

For each of the two examples, the training data taken by the learning
algorithm are a discrete time sequence generated by solving the EL
equation of an exact continuous Lagrangian. In 1D, the EL equation
is an Ordinary Differential Equation (ODE) in time. Only the training
sequence is visible to the learning and serving algorithms, and the
EL equation and the continuous Lagrangian are not. After learning
the discrete Lagrangian from the training data, the algorithm serves
it by predicting new dynamic sequences $\psi_{i}$ for different initial
conditions. The predictions are compared with accurate numerical solutions
of the EL equation.

Before presenting the numerical results, I briefly describe how the
algorithms are implemented. To learn $L_{d}(\psi_{i},\psi_{i+1})$,
a neural network can be set up. Since it has only two inputs and one
output, a deep network may not be necessary. For these two specific
examples, the functional approximation for $L_{d}(\psi_{i},\psi_{i+1})$
is implemented using polynomials in terms of $s\equiv\psi_{i}+\psi_{i+1}$
and $d\equiv\psi_{i+1}-\psi_{i}$, i.e., 
\begin{equation}
L_{d}(\psi_{i},\psi_{i+1})=\sum_{p=0}^{P}\sum_{q=0}^{Q}a_{pq}d^{p}s^{q}\,,
\end{equation}
where $a_{pq}$ are trainable parameters. For these two examples,
I choose $(P,Q)=(4,8)$, and the total number of trainable parameters
are $45$. For high-dimensional or vector discrete field theories,
such as the Kepler problem in Sec.\,\ref{sec:Kepler}, deep neural
networks are probably more effective.
\begin{example}
The training data are plotted in Fig.\,\ref{fig:ex1traning} using
empty square markers. It is a time sequence $\bar{\psi}_{i}$ $(i=0,...,50)$
generated by the nonlinear ODE
\begin{equation}
2\left(\sin\psi+1\right)\psi^{\prime\prime}+\left(\psi^{\prime}\right)^{2}\cos\psi+\frac{\pi^{2}}{200}\psi=0\label{ex1ODE}
\end{equation}
with initial conditions $\psi(t=0)=1.2$ and $\psi^{\prime}(t=0)=0.$
Here $\psi^{\prime}$ denote $d\psi/dt$. The Lagrangian density for
the system is 
\begin{equation}
L(\psi,\psi^{\prime})=\left(1+\sin\psi\right)\left(\psi^{\prime}\right)^{2}-\frac{\pi^{2}}{400}\psi^{2}\,.\label{ex1L}
\end{equation}
The optimizer for training the discrete Lagrangian density $L_{d}$
is Newton's algorithm with step-lengths reduced according to the amplitude
of loss function. The discrete Lagrangian density $L_{d}$ is trained
until the loss function on $\bar{\psi}$ is less than $10^{-7}$,
then it is served. Plotted in Fig.\,\ref{fig:ex1traning} using solid
circle markers are the predicted time sequence $\psi_{i}$ using the
initial conditions of the training data, i.e., $\psi_{0}=\bar{\psi}_{0}$
and $\psi_{1}=\bar{\psi}_{1}.$ The predicted sequence $\psi_{i}$
and the training sequence $\bar{\psi}_{i}$ are barely distinguishable
in the figure.

\begin{figure}[ptb]
\begin{centering}
\includegraphics[width=4in]{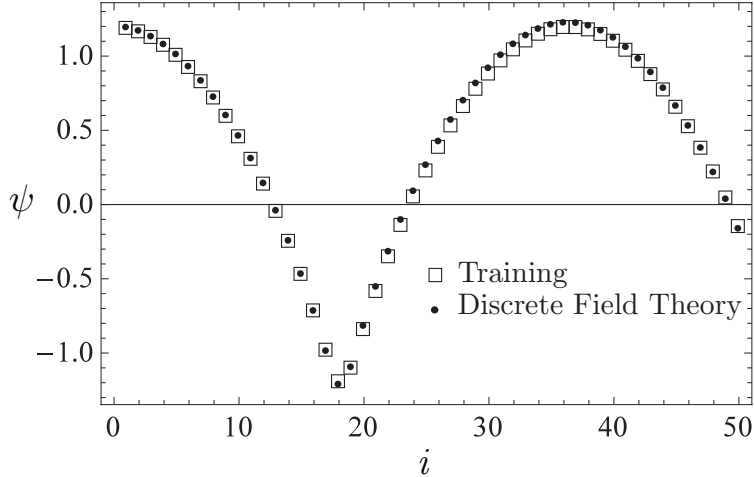}
\par\end{centering}
\caption{The predicted sequence $\psi_{i}$ (solid circles) from the learned
discrete field theory and the training sequence $\bar{\psi}_{i}$
(empty squares) are barely distinguishable in the figure. The discrete
Lagrangian is trained until the loss function $F(\overline{\psi})$
is less than $10^{-7}$.}

\label{fig:ex1traning}
\end{figure}

The learned discrete field theory is then served with two sets of
new initial conditions, and the predicted time sequences are plotted
using solid circle markers in Fig.\,\ref{fig:ex1test1} and Fig.\,\ref{fig:ex1test2}
against the time sequences solved for from the nonlinear ODE (\ref{ex1ODE}).
The predicted sequence in Fig.\,\ref{fig:ex1test1} starts at $\psi_{0}=-0.6$,
and its dynamic characteristics is significantly different from that
of the sequence in Fig.\,\ref{fig:ex1traning}. The predicted sequence
in Fig.\,\ref{fig:ex1test2} starts at a much smaller amplitude,
i.e., $\psi_{0}=0.1$, and shows the behavior of linear oscillation,
in contrast with the strong nonlinearity of the sequence in Fig.\,\ref{fig:ex1traning}
and the mild nonlinearity of the sequence in Fig.\,\ref{fig:ex1test1}.
The agreement between the predictions of the learned discrete field
theory and the accurate solutions of the nonlinear ODE (\ref{ex1ODE})
is satisfactory. These numerical results demonstrate that the proposed
algorithms for machine learning and serving of discrete field theories
are effective in terms of capturing the fundamental structure and
predicting the complicated dynamical behavior of the physical system.
\end{example}

\begin{figure}[ptb]
\begin{centering}
\includegraphics[width=4in]{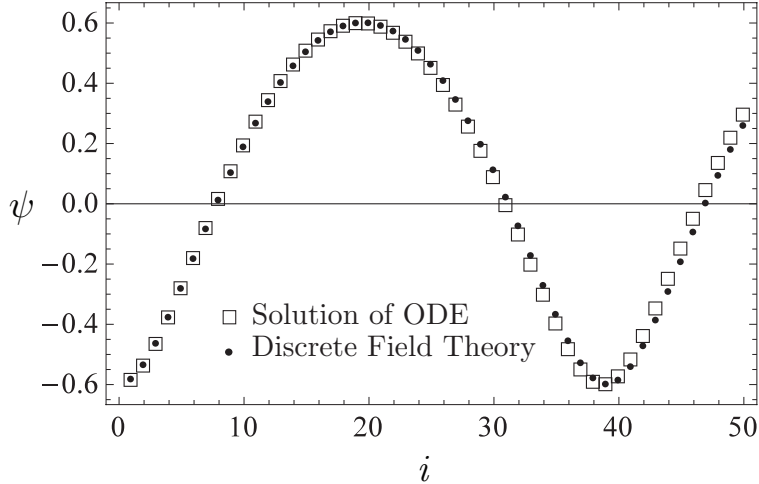}
\par\end{centering}
\caption{The predicted time sequence (solid circles) agrees with the time sequence
(empty squares) accurately solved for from the nonlinear ODE (\ref{ex1ODE}).
The dynamics starts at $\psi_{0}=-0.6$, and its characteristics is
significantly different from that of the time sequence in Fig.\,\ref{fig:ex1traning}.}

\label{fig:ex1test1}
\end{figure}

\begin{figure}[ptb]
\begin{centering}
\includegraphics[width=4in]{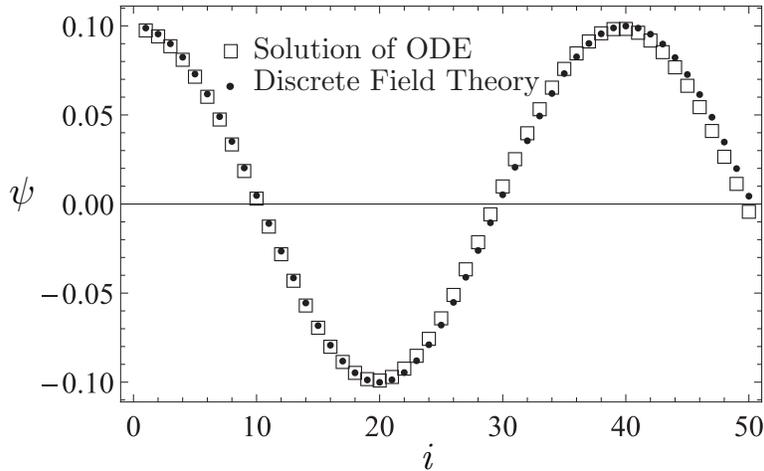}
\par\end{centering}
\caption{Starting at a much smaller amplitude, i.e., $\psi_{0}=0.1$, the predicted
sequence (solid circles) shows the behavior of linear oscillation,
in contrast with the strong nonlinearity of the sequence in Fig.\,\ref{fig:ex1traning}
and the mild nonlinearity of the sequence in Fig.\,\ref{fig:ex1test1}.
The predicted time sequence agrees with the time sequence (empty squares)
accurately solved for from the nonlinear ODE (\ref{ex1ODE}).}

\label{fig:ex1test2}
\end{figure}

\begin{figure}[ptb]
\begin{centering}
\includegraphics[width=4in]{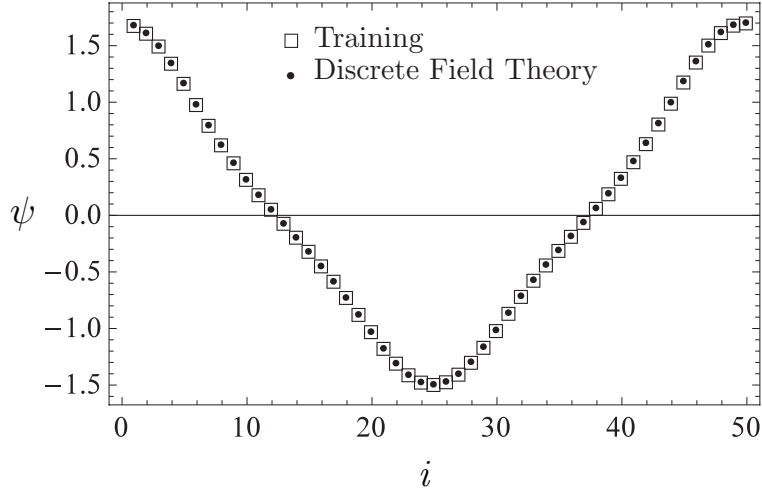}
\par\end{centering}
\caption{The training sequence (empty squares) represents a nonlinear oscillation
in potential wall between $\psi=\pm1.6$ in Fig.\,\ref{fig:ex2V}.
The trained discrete Lagrangian density $L_{d}$ is accepted when
the loss function $F(\overline{\psi})$ on the training sequence is
less than $10^{-7}$. The predicted sequence (solid circles) from
the learned discrete field theory agrees very well with the training
sequence.}

\label{fig:ex2traing}
\end{figure}

\begin{figure}[ptb]
\begin{centering}
\includegraphics[width=4in]{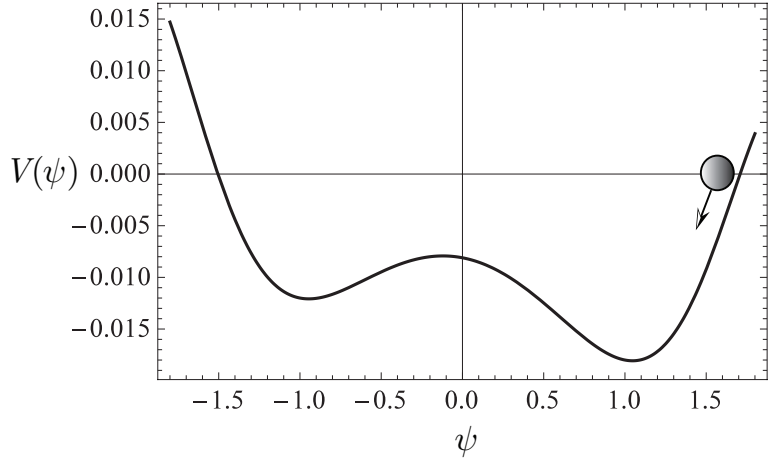}
\par\end{centering}
\caption{The training sequence  in Fig.\,\ref{fig:ex2traing} represents a
nonlinear oscillation in the large potential wall between $\psi=\pm1.6$.
There are two small potential wells secondary to the large potential
well, one on the left between between $\psi=-1.3$ and $\psi=-0.1$,
and one on the right between $\psi=-0.1$ and $\psi=1.5$.}

\label{fig:ex2V}
\end{figure}

\begin{figure}[ptb]
\begin{centering}
\includegraphics[width=4in]{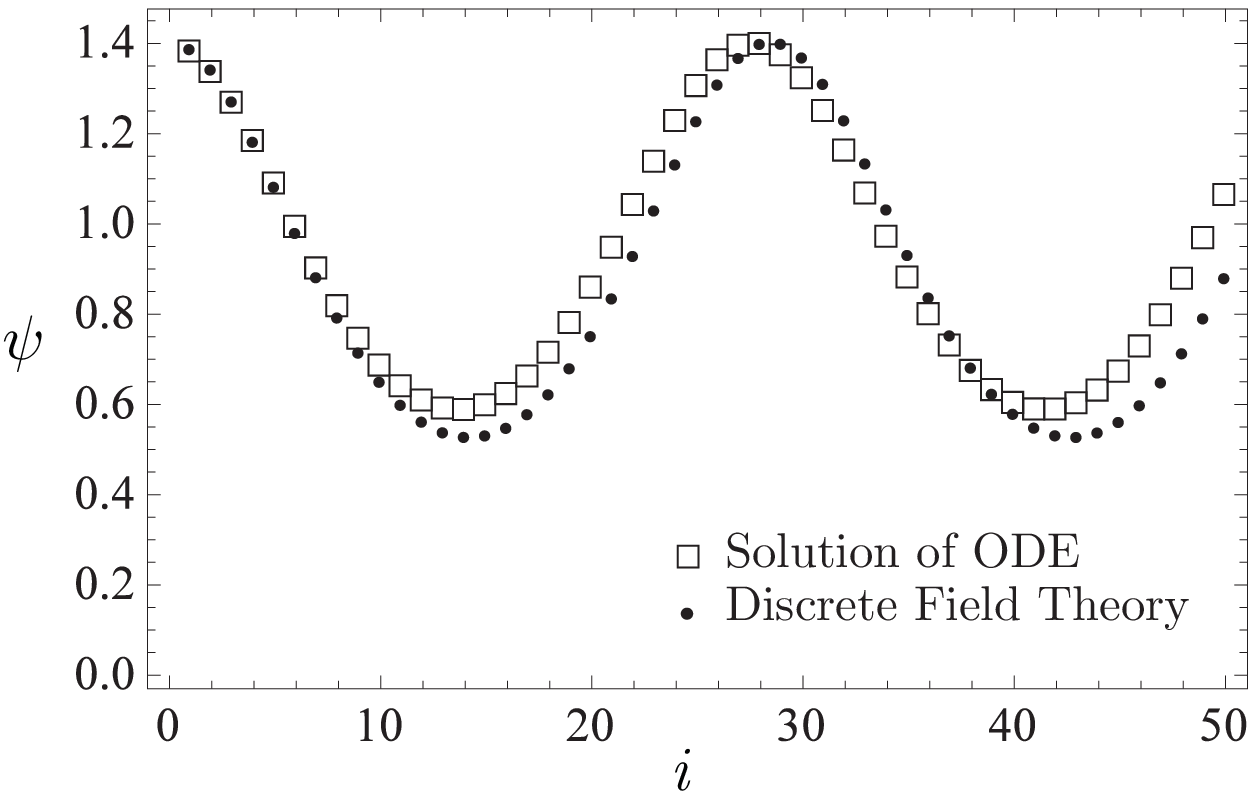}
\par\end{centering}
\caption{The learned discrete field theory correctly predicts a nonlinear oscillation
in the small potential well between $\psi=-0.1$ and $\psi=1.5$ on
the right of Fig.\,\ref{fig:ex2V}. The predicted sequence (solid
circles) agrees with the accurate solution (empty squares) of the
nonlinear ODE (\ref{ex2ODE}).}

\label{fig:ex2test1}
\end{figure}

\begin{figure}[ptb]
\begin{centering}
\includegraphics[width=4in]{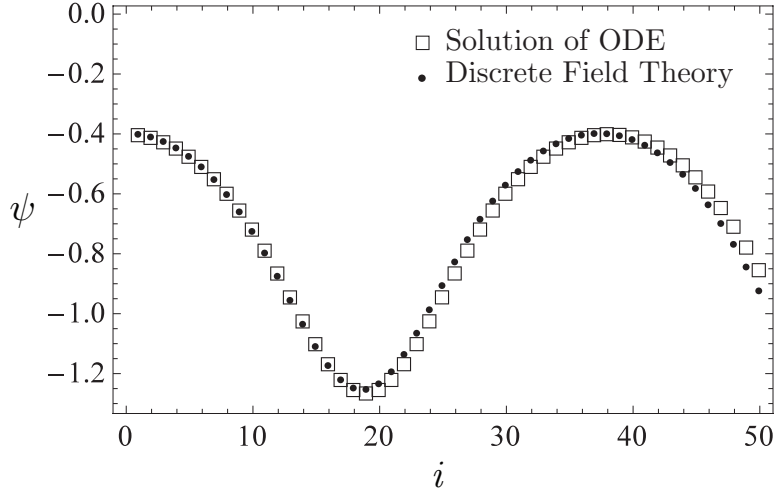}
\par\end{centering}
\caption{The learned discrete field theory correctly predicts a nonlinear oscillation
in the small potential well between $\psi=-1.3$ and $\psi=-0.1$
on the left of Fig.\,\ref{fig:ex2V}. The predicted sequence (solid
circles) agrees with the accurate solution (empty squares) of the
nonlinear ODE (\ref{ex2ODE}).}

\label{fig:ex2test2}
\end{figure}

\begin{example}
The training data are plotted in Fig.\,\ref{fig:ex2traing} using
empty square markers. It is a time sequence $\bar{\psi}_{i}$ $(i=0,...,50)$
generated by the nonlinear ODE
\begin{equation}
\psi^{\prime\prime}-0.03\left[\sin\left(1-\psi^{2}\right)\psi+0.1\right]=0\label{ex2ODE}
\end{equation}
with initial conditions $\psi(t=0)=1.7$ and $\psi^{\prime}(t=0)=0.$
The Lagrangian for the system is 
\begin{align}
L(\psi,\psi^{\prime}) & =\frac{1}{2}\left(\psi^{\prime}\right)^{2}-V(\psi)\,,\label{ex2L}\\
V(\psi) & =-0.015\left[\cos\left(\psi^{2}-1\right)+0.2\psi\right]\,,\label{ex2V}
\end{align}
where $V(\psi)$ is a nonlinear potential plotted in Fig.\,\ref{fig:ex2V}.
The training sequence represents a nonlinear oscillation in the potential
well between $\psi=\pm1.6$. The trained discrete Lagrangian density
$L_{d}$ is accepted when the loss function $F(\overline{\psi})$
on the training sequence is less than $10^{-7}$. The predicted sequence
$\psi_{i}$ (solid circles in Fig.\,\ref{fig:ex2traing}) by the
serving algorithm from the learned discrete field theory agrees very
well with the training sequence $\bar{\psi}_{i}$.

The learned discrete field theory predicts two very different types
of dynamical sequences shown in Fig.\,\ref{fig:ex2test1} and Fig.\,\ref{fig:ex2test2}.
The predicted sequences are plotted using solid circle markers and
the sequences accurately solved for from the nonlinear ODE (\ref{ex2ODE})
are plotted using empty square markers. The sequence predicted in
Fig.\,\ref{fig:ex2test1} is a nonlinear oscillation in the small
potential well between $\psi=-0.1$ and $\psi=1.5$ on the right of
Fig.\,\ref{fig:ex2V}, and the sequence predicted in Fig.\,\ref{fig:ex2test2}
is a nonlinear oscillation in the small potential well between $\psi=-1.3$
and $\psi=-0.1$ on the left. For both cases, the predictions of the
learned discrete field theory agree with the accurate solutions. Observe
that in Fig.\,\ref{fig:ex2V} the two small potential wells are secondary
to the large potential wall between $\psi=\pm1.6$. In Fig.\,\ref{fig:ex2traing}
the small-scale fluctuations in the training sequence, which is a
nonlinear oscillation in the large potential well, encode the structures
of the small potential wells. The training algorithm is able to diagnose
and record these fine structures in the learned discrete Lagrangian
density, and the serving algorithm correctly predicts the secondary
dynamics due to them.
\end{example}

\section{Kepler problem \label{sec:Kepler}}

\begin{figure}[t]
\begin{centering}
\includegraphics[width=4in]{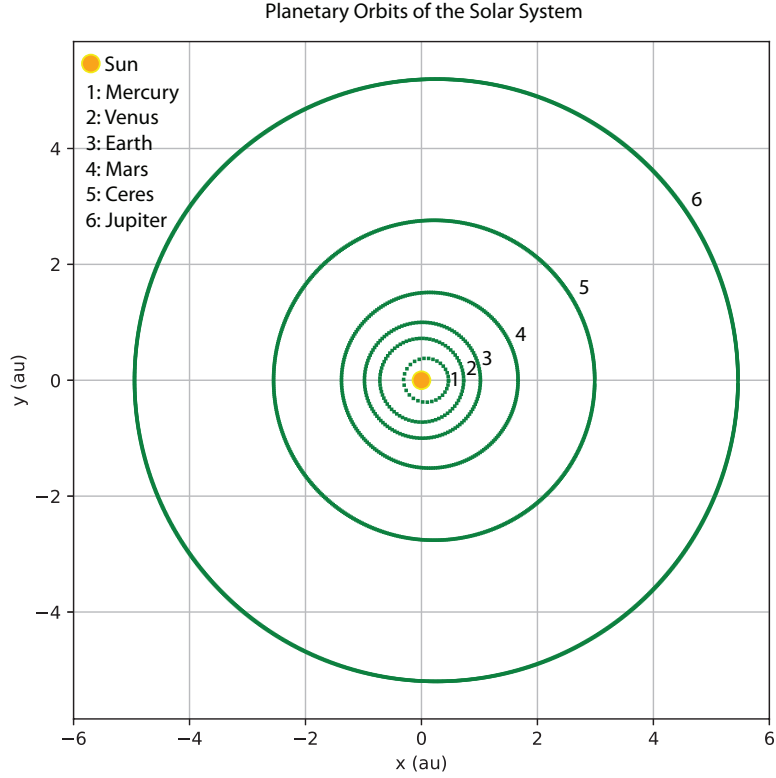}
\par\end{centering}
\caption{Orbits of the Mercury, Venus, Earth, Mars, Ceres and Jupiter generated
by solving Newton's equation of motion for a planet in the gravity
field of the sun according to Newton's law of universal gravitation.
These orbits are the training data for the discrete field theory.}

\label{fig:training orbits}
\end{figure}

In this section, to further demonstrate the effectiveness of the method
developed, I apply it to the Kepler problem, which is concerned with
dynamics of planetary orbits in the solar system. Let turn the clock
back to 1601, when Kepler inherited the observational data of planetary
orbits meticulously collected by his mentor Tycho Brahe. It took Kepler
5 years to discover his first and second laws of planetary motion,
and another 78 years before Newton solved the Kepler problem using
his laws of motion and universal gravitation \citep{Newton2008}.
Assume that we have a set of data similar to that of Kepler, as displayed
in Fig.\,\ref{fig:training orbits}. For simplicity, the data are
the orbits of the Mercury, Venus, Earth, Mars, Ceres and Jupiter generated
by solving Newton's equation of motion for a planet in the gravity
field of the Sun according to Newton's law of universal gravitation.
The spatial and temporal normalization scale-lengths are 1 a.u. and
58.14 days, respectively, and the time-steps of the orbital data is
0.05.

\begin{figure}
\begin{centering}
\includegraphics[width=4in]{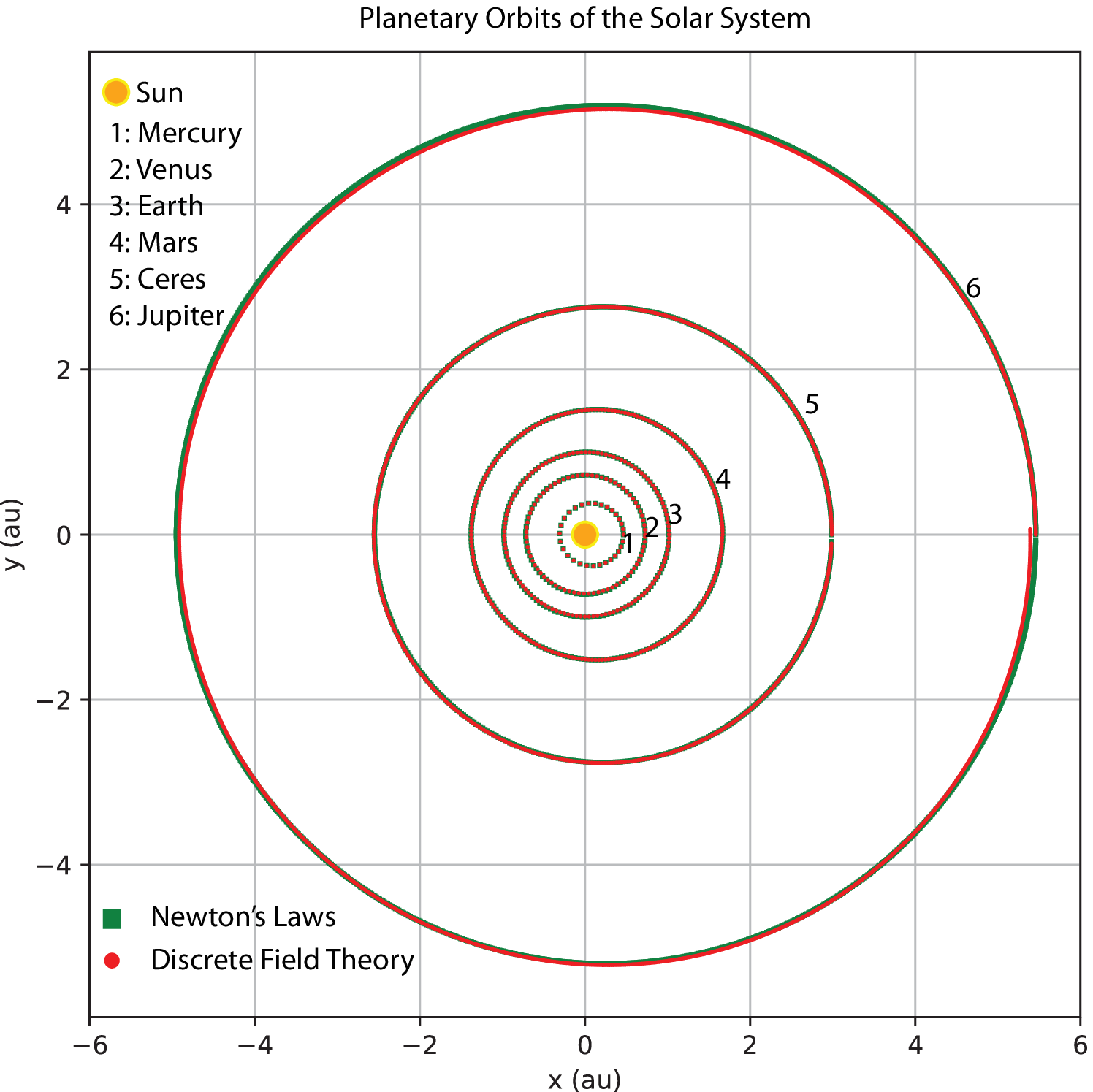}
\par\end{centering}
\caption{Orbits of the Mercury, Venus, Earth, Mars, Ceres and Jupiter. The
orbits indicated by red markers are generated by the learned discrete
field theory. The orbits indicated by green markers are the training
orbits from Fig.\,\ref{fig:training orbits}.}

\label{fig:learned orbits}
\end{figure}

My goal here is not to rediscover Kepler's laws of planetary motion
or Newton's laws of motion and universal gravitation by machine learning.
Instead, I train a discrete field theory from the orbits displayed
in Fig.\,\ref{fig:training orbits} and then serve it to predict
new planetary orbits. For this case, the discrete field theory is
about a 2D vector field defined on the time grid. Denote the field
as $\psi_{i}=(x_{i},y_{i})$, where $i$ is the index for the time
grid, and $x_{i}$ and $y_{i}$ are the 2D coordinates of a planet
in the solar system. In terms of the discrete field, the discrete
Lagrangian density $L_{d}$ is a function on $\mathrm{R}^{4},$
\begin{equation}
L_{d}(\psi_{i},\psi_{i+1})=L_{d}(x_{i},y_{i},x_{i+1},y_{i+1})\,,\label{eq:Ld4}
\end{equation}
 and the DEL is a vector equation with two components,
\begin{gather}
EL_{x_{i}}=\frac{\partial L_{d}(x_{i-1},y_{i-1},x_{i},y_{i})}{\partial x_{i}}+\frac{\partial L_{d}(x_{i},y_{i},x_{i+1},y_{i+1})}{\partial x_{i}}=0\,,\label{Elx}\\
EL_{y_{i}}=\frac{\partial L_{d}(x_{i-1},y_{i-1},x_{i},y_{i})}{\partial y_{i}}+\frac{\partial L_{d}(x_{i},y_{i},x_{i+1},y_{i+1})}{\partial y_{i}}=0\,.\label{Ely}
\end{gather}
The loss function on a data set $\psi=(x,y)$ is 
\[
F(x,y)=\frac{1}{I}\sum_{i=1}^{I-1}\left[EL_{x_{i}}(x,y)^{2}+EL_{y_{i}}(x,y)^{2}\right]\,.
\]
Akin to the situation in Sec.\,\ref{sec:Examples}, the serving algorithms
preserves exactly an discrete symplectic form defined by 
\begin{gather}
\Omega(x_{i},y_{i},x_{i+1},y_{i+1})=d\theta\,,\\
\theta=\frac{\partial L_{d}(x_{i},y_{i},x_{i+1},y_{i+1})}{\partial x_{i+1}}dx_{i+1}+\frac{\partial L_{d}(x_{i},y_{i},x_{i+1},y_{i+1})}{\partial y_{i+1}}dy_{i+1}\,.
\end{gather}

\begin{figure}
\begin{centering}
\includegraphics[width=4in]{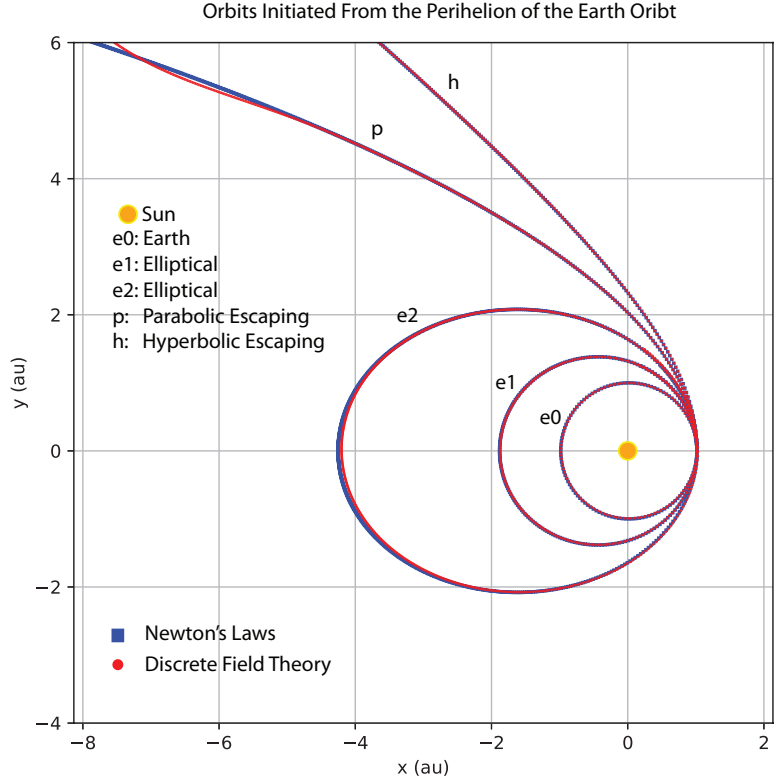}
\par\end{centering}
\caption{Orbits initiated from the Perihelion of the Earth orbit with initial
velocities $v_{p}=0.98,1.13,1.26,1.40,\text{ and }1.50.$ Orbit e0
is the Earth orbit. Orbits e1 and e2 are elliptical, and Orbit p is
the parabolic escaping orbit and Orbit h is the hyperbolic escaping
orbit. Red markers are the predictions of the trained discrete field
theory, and blue markers are solutions according to Newton's laws
of motion and universal gravitation.}

\label{fig: Earth}
\end{figure}

To model the discrete Lagrangian density $L_{d}(x_{i},y_{i},x_{i+1},y_{i+1})$,
I use a fully connected neural network with two hidden layers, each
of which has 40 neurons with the sigmoid activation function. The
network is randomly initialized with a normal distribution, and then
trained by the Adam optimizer \citep{Kingma2014} until the averaged
loss on a single time grid-point is reduced by a factor of $10^{5}$
relative to its initial value. Starting from the same initial conditions
as the training orbits, the serving algorithm of the trained discrete
field theory predicts the orbits plotted using red markers in Fig.\,\ref{fig:learned orbits}
against the training orbits indicated by green markers. The agreement
between the predicted and training orbits shown in the figure validates
the discrete field theory learned. To serve it for the purpose of
predicting new orbits, let's consider the scenario of launching a
device at the Perihelion of the Earth orbit with an orbital velocity
$v_{p}$ larger than that of the Earth. Four such orbits, labeled
by e1, e2, h, and p with $v_{p}=1.13,$ $1.26,$ $1.40,$ and $1.50,$
are plotted in Fig.\,\ref{fig: Earth} along with the orbit of the
Earth, which is the inner most ellipse labeled by e0 with $v_{p}=0.98$.
Orbits plotted using red markers are predictions of the trained discrete
field theory, and orbits plotted using blue markers are solutions
according to Newton's laws of motion and gravitation. The agreement
is excellent. Orbits e1 and e2 are elliptical, and Orbit p is the
parabolic escaping orbit and Orbit h is the hyperbolic escaping orbit.

\begin{figure}
\begin{centering}
\includegraphics[width=4in]{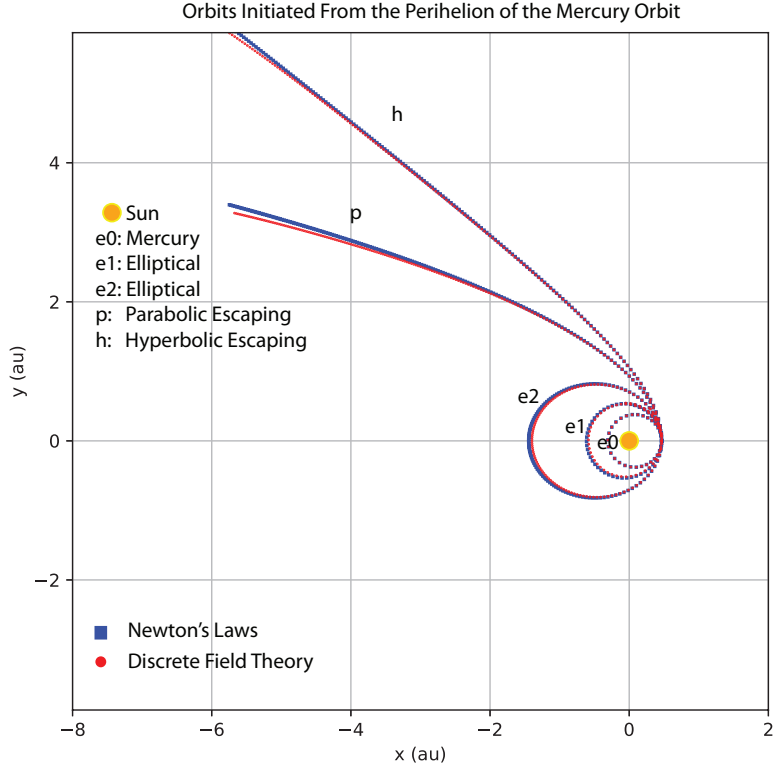}
\par\end{centering}
\caption{Orbits initiated from the Perihelion of the Mercury orbit with initial
velocities $v_{p}=1.30,1.56,1.80,2.07,\text{ and }2.20.$ Plotting
markers and labels are similar to those in Fig.\,\ref{fig: Earth}.}

\label{fig:Mercury}
\end{figure}

Similar study is carried out for the orbits initiated from the Perihelion
of the Mercury orbit with an orbital velocities $v_{p}$ larger than
that of the Mercury. Four such orbits are shown in Fig.\,\ref{fig:Mercury}
using the same plotting markers and labels as in Fig.\,\ref{fig: Earth}.
The inner most elliptical orbit is that of the Mercury with $v_{p}=1.30$.
The orbit velocities at the Perihelion of the other four orbits are
$v_{p}=1.56,$ $1.80$, $2.07,$ $2.20.$ Again, the the predictions
of the trained discrete field theory agree very well with those of
Newton's laws.

It is remarkable that the trained discrete field theory correctly
predicts the parabolic and hyperbolic escaping orbits, even though
the training orbits are all elliptical, see Figs.\,\ref{fig:training orbits}
and \ref{fig:learned orbits}. Historically, Kepler argued that escaping
orbits and elliptical orbits are governed by different laws. It was
Newton who discovered or ``learned'' the $1/r$ dependency of the
gravitational field from Kepler's laws of planetary motion and Tycho
Brahe's data, and unified the elliptical orbits and escaping orbits
under the same law of physics. Most of the studies physicists have
been doing since then is applying Newton's methodology to other physical
phenomena. The results displayed in Figs.\,\ref{fig: Earth} and
\ref{fig:Mercury} show that the machine learning and serving algorithms
solve the Kepler problem in terms of correctly prediction planetary
orbits without knowing or learning Newton's laws of motion and universal
gravitation.

To complete this section, a few footnotes are in order. (i) There
exist small discrepancies between the predictions from the learned
discrete field theory and Newton's laws in Figs.\,\ref{fig: Earth}
and \ref{fig:Mercury} when $r=\sqrt{x^{2}+y^{2}}\gtrsim7$. This
is because no training orbit in this domain was provided to the learning
algorithm. The orbits predicted there are thus less accurate. (ii)
The study presented is meant to be a proof of principle. Practical
factors, such as three-body effects, are not included. Nevertheless,
the method itself is robust against variations of the governing laws
of physics, because the method does not require any knowledge of the
laws of physics other than the fundamental assumption that the governing
laws are field theories. In particular, the learning and serving algorithms
for planetary orbits described above do not assume or make use of
Newton's equation of motion and Newton's law of universal gravitation.
Therefore, when the effects of special relativity or general relativity
are important, the algorithms are valid without modification. Further
study will be reported in the future.

\section{Conclusions and discussion}

In this paper, a method for machine learning and serving of discrete
field theories in physics is developed. The learning algorithm trains
a discrete field theory from a set of observational data of the field
on a spacetime lattice, and the serving algorithm employs the learned
discrete field theory to predict new observations of the field for
given new boundary and initial conditions.

The algorithm does not attempt to capture statistical properties of
the training data, nor does it try to discover differential equations
that govern the training data. Instead, it learns a discrete field
theory that underpins the observed field. Because the learned field
theory is discrete, it overcomes the difficulties associated with
the learning of continuous theories. Compared with continuous field
theories, discrete field theories can be served more easily and with
improved long-term accuracy and fidelity. The serving algorithm of
discrete field theories belongs to the family of structure-preserving
geometric algorithms \citep{Feng1985,Sanz-Serna1994,marsden1998multisymplectic,marsden2001discrete,hairer2006geometric,Qin2008VI-PRL,qin2009variational,li2011variational,Squire4748,squire2012geometric,squire2012gauge,Xiao2013,zhang2014canonicalization,zhou2014variational,he2015Hamiltonian,xiao2015explicit,xiao2015variational,ellison2015development,qin2016canonical,He16-092108,xiao2016explicit,zhang2016explicit,Wang2016,xiao2017local,burby2017finite,chen2017canonical,zhou2017explicit,He2017symplectic,Burby2017,kraus2017gempic,Xiao2018b,Ellison2018,Xiao2019Maxwell,Xiao2019,Xiao2019a,Glasser2019b,Shi2019,Xiao2019-comment},
which have been proven to be superior to the conventional algorithms
based on discretization of differential equations. The demonstrated
advantages of discrete field theories relative to continuous theories
in terms of machine learning compatibility are consistent with Bostrom's
simulation hypothesis. The synergy between artificial intelligence
and the concept of discrete universe may bring pleasant surprises.

Finally, I should emphasize that no machine learning algorithm is
meaningful or effective without presumptions. The algorithms developed
here certainly do not apply to any given set of data. The data relevant
to the present study are assumed to be observations of physical fields
in the spacetime governed by field theories.

However, the existence of a governing field theory is the only physical
assumption required. Laws of physics in specific forms, such as Newton's
laws of motion and gravity, special relativity and general relativity,
and Schrödinger's equation, are not needed for the machine learning
and serving algorithms of discrete field theories to be effective
in terms of correctly predicting observations.
\begin{acknowledgments}
This research was supported by the U.S. Department of Energy (DE-AC02-09CH11466).
I thank Alexander S. Glasser, Yichen Fu, Michael Churchill, George
Wilkie, and Nick McGreivy for fruitful discussions.
\end{acknowledgments}

\bibliographystyle{apsrev4-1}
\bibliography{MLDFT2}

\end{document}